\documentclass[preprint,showpacs,amsmath,amssymb]{revtex4}
\usepackage{mathrsfs}

\usepackage{graphicx,color}
\usepackage{dcolumn}
\usepackage{bm}
\usepackage{CJK}

\begin{document}
\begin{CJK*}{GBK}{song}
%-------------------------------------------------------------------------------------------------------

\title{Radial basis function approach in nuclear mass predictions}

\author{Z. M. Niu$^1$}\email{zmniu@ahu.edu.cn}
\author{Z. L. Zhu$^1$}
\author{Y. F. Niu$^2$}
\author{B. H. Sun$^3$}\email{bhsun@buaa.edu.cn}
\author{T. H. Heng$^1$}\email{hength@ahu.edu.cn}
\author{J. Y. Guo$^1$}\email{jianyou@ahu.edu.cn}

\affiliation{$^1$School of Physics and Material Science, Anhui University,
             Hefei 230039, China}
\affiliation{$^2$Institute of Fluid Physics, China Academy of Engineering
             Physics, Mianyang 621900, China}
\affiliation{$^3$School of Physics and Nuclear Energy Engineering, Beihang University,
             Beijing 100191, China}

\date{\today}

%-------------------------------------------------------------------------------------------------------
\begin{abstract}
The radial basis function (RBF) approach is applied in predicting nuclear masses for $8$ widely used nuclear mass models, ranging from macroscopic-microscopic to microscopic types. A significantly improved accuracy in computing nuclear masses is obtained, and the corresponding rms deviations with respect to the known masses is reduced by up to $78\%$. Moreover, strong correlations are found between a target nucleus and the reference nuclei within about three unit in distance, which play critical roles in improving nuclear mass predictions. Based on the latest Weizs\"{a}cker-Skyrme mass model, the RBF approach can achieve an accuracy comparable with the extrapolation method used in atomic mass evaluation. In addition, the necessity of new high-precision experimental data to improve the mass predictions with the RBF approach is emphasized as well.
\end{abstract}

\pacs{21.10.Dr, 21.60.-n} \maketitle
%21.10.Dr Binding energies and masses
%21.60.-n Nuclear structure models and methods
%-------------------------------------------------------------------------------------------------------
\section{Introduction}
Nuclear mass plays an important role not only in studying the knowledge of nuclear structure~\cite{Lunney2003RMP}, but also in understanding the origin of elements in the universe~\cite{Burbidge1957RMP, Arcones2012PRL}. With the construction and upgrade of radioactive ion beam facilities, the measurements of nuclear masses have made great progress in recent years. During the last decade, hundreds of nuclear masses were measured for the first time or with higher precisions~\cite{Audi2012CPC}.

The astrophysical calculations involve thousands of nuclei far from $\beta$-stability line. However, most of these nuclei are still beyond the experimental reach. One could use the local mass relations such as the Garvey-Kelson (GK) relations~\cite{Garvey1966PRL, Garvey1969RMP} and the residual proton-neutron interactions~\cite{Zhang1989PLB, Fu2010PRC, Jiang2010PRC, Jiang2012PRC} to predict unknown masses. However, the intrinsic error grows rapidly when the local mass relations are used to predict the nuclear masses in an iterative way~\cite{Fu2011PRC, Morales2009NPA}. Therefore, the theoretical predictions for nuclear masses are inevitable to astrophysical calculations. The early theoretical studies of nuclear masses are mainly macroscopic models, such as the famous Weizs\"{a}cker mass formula~\cite{Weizsacker1935ZP}. It is known that this kind of mass model neglects the microscopic effects, and hence shows systematic deviations for the nuclei near the shell closure or those with large deformations. In order to better describe the nuclear ground-state properties, the macroscopic-microscopic and microscopic theoretical models are developed for mass predictions.

By including the microscopic correction energy to the macroscopic mass formula, the macroscopic-microscopic mass model can well take into account the important microscopic corrections. During the past decades, a number of macroscopic-microscopic mass models have been developed, such as the finite-range droplet model (FRDM)~\cite{Moller1995ADNDT}, the extended Thomas-Fermi plus Strutinsky integral (ETFSI)~\cite{Goriely2000AIP}, and the Koura-Tachibana-Uno-Yamada (KTUY)~\cite{Koura2005PTP}. These macroscopic-microscopic mass models have similar accuracy for mass prediction and their the root-mean-square (rms) deviation with respect to data in the atomic mass evaluation of 2012 (AME12)~\cite{Audi2012CPC} is about $0.7$ MeV. Guiding by the Skyrme energy density functional, a semiempirical nuclear mass formula, the Weizs\"{a}cker-Skyrme (WS) model, was proposed based on the macroscopic-microscopic method~\cite{Wang2010PRCa, Wang2010PRCb, Liu2011PRC}. For the latest version of WS model (WS3)~\cite{Liu2011PRC}, the rms deviation with respect to $2353$ known nuclear masses in AME12 is significantly reduced to $0.335$ MeV.

On the other hand, great progress has been achieved for microscopic mass models with the rapid development of the computer technology in the new century. Based on the Hartree-Fock-Bogoliubov (HFB) theory with Skyrme or Gogny force, a series of microscopic mass models have been proposed with the accuracy comparable with the traditional macroscopic-microscopic mass models~\cite{Goriely2009PRLa, Goriely2009PRLb, Goriely2010PRC}. Apart from the non-relativistic microscopic model, the relativistic mean-field (RMF) model has also received wide attention due to many successes achieved in describing lots of nuclear phenomena~\cite{Meng2006PPNP, Vretenar2005PRp, Zhao2010PRC, Hua2012SCMPA, Zhou2010PRC, Long2010PRC, Niu2013PRC, Mei2012PRC, Liang2008PRL, Niu2009PLB, Vretenar2012PRC, Liang2013PRC, Niu2013PRCR} as well as successful applications in astrophysics~\cite{Sun2008PRC, Sun2008CPL, Niu2009PRC, Niu2011PRC, Niu2011IJMPE, Meng2011SCSG, Zhang2012APS, Xu2013PRC, Niu2013PLB}. A systematic study of the ground-state properties for all nuclei from the proton drip line to the neutron drip line with $Z, N\geqslant 8$ and $Z\leqslant 100$ was performed for such model several years ago, and the rms deviation with respect to known masses is about $2$ MeV~\cite{Geng2005PTP}. However, it should be noted that the effective interaction of this RMF mass model was only optimized with the properties of a few selected nuclei. By carefully adjusting the effective interaction of RMF model with the properties of more selected nuclei, the deviation can be remarkably reduced. For the $575$ even-even nuclei with $8\leqslant Z\leqslant 108$, the rms deviation with respect to known masses in atomic mass evaluation of 2003 (AME03) is reduced to $1.24$ MeV for the effective interaction PC-PK1~\cite{Zhang2013arXiv}. Moreover, the PC-PK1 predictions well reproduce the new and accurate mass measurements from Sn to Pa~\cite{Chen2012NPA} with the rms deviation of $0.859$ MeV~\cite{Zhao2012PRC}, and also successfully describe the Coulomb displacement energies between mirror nuclei~\cite{Sun2011SCPMA}. In addition, inspired by the shell model, the Duflo-Zuker (DZ) mass model~\cite{Duflo1995PRC, Zuker2008RMF} has made considerable success in describing nuclear masses with accuracy of about $0.5$ MeV.

Although these theoretical models can well reproduce the experimental data, there are still large deviations among the mass predictions of different models, even in the region close to known masses. A number of investigations on the accuracy and predictive power of these nuclear mass models have been made so far in the literatures, e.g. Refs.~\cite{Lunney2003RMP, Geng2005PTP, Temis2008NPA, Litvinov2012IJMPE, Sobiczewski2013PST}. To further improve the accuracy of nuclear mass model, the image reconstruction techniques based on the Fourier transform is applied to the nuclear mass models and significantly reduces the rms deviation to the known masses with the CLEAN algorithm~\cite{Morales2010PRC}. Later on, the radial basis function (RBF) approach was developed to improve the mass predictions of several theoretical models~\cite{Wang2011PRC}. Comparing with the CLEAN reconstruction, the RBF approach more effectively reduces the rms deviations with respect to the masses first appearing in AME03~\cite{Wang2011PRC}.

To improve the mass prediction of a nucleus, thousands of nuclei with known masses are involved in the RBF approach~\cite{Wang2011PRC}. However, do all the nuclei involved play effective roles in the improvement of mass prediction for this nucleus? What are the key nuclei that have to be included in the RBF approach? In other words, how far away from the measured region of nuclear mass could we predict with satisfactory accuracy in the RBF approach? These questions were not addressed in previous investigations~\cite{Wang2011PRC}. Therefore, it is interesting to investigate the mass correlations between a certain nucleus and those nuclei involved in the RBF approach, and hence to evaluate the predictive power of the RBF approach.

In this work, we will carefully evaluate the predictive power of the RBF approach based on $8$ widely used nuclear mass models, ranging from macroscopic-microscopic to microscopic types. Special attention will be paid to the mass correlations among various nuclei. The paper is organized as follows. In Sec. II, a brief introduction to the RBF approach including numerical details is given. In Sec. III, the mass correlations are first carefully investigated, and then the predictive power of the RBF approach based on different mass models will be evaluated. Finally, the summary is presented in Sec. IV.

\section{Radial basis function approach and numerical details}
The RBF approach has been widely applied in surface reconstruction and its solution is written as
\begin{eqnarray}\label{Eq:Sx}
  S(x) = \sum_{i=1}^m \phi(\|x-x_i\|) \omega_i,
\end{eqnarray}
where $x_i$ denotes the point from measurement, $\omega_i$ is the weight of center $x_i$,
$\phi$ is the radial basis function, $\|x-x_i\|$ is the Euclidean norm, and $m$ is the number
of the data to be fitted. Given $m$ samples $(x_i, d_i)$, one wishes to reconstruct the smooth function $S(x)$ with $S(x_i)=d_i$, i.e.,
\begin{eqnarray}\label{Eq:Dx}
\left(
  \begin{array}{c}
    d_1 \\
    d_2 \\
    ... \\
    d_m \\
  \end{array}
\right)
=
\left(
  \begin{array}{cccc}
    \phi_{11} & \phi_{12} & ... & \phi_{1m} \\
    \phi_{21} & \phi_{22} & ... & \phi_{2m} \\
    ...       & ...       & ... & ... \\
    \phi_{m1} & \phi_{m2} & ... & \phi_{mm} \\
  \end{array}
\right)
\left(
  \begin{array}{c}
    \omega_1 \\
    \omega_2 \\
    ... \\
    \omega_m \\
  \end{array}
\right),
\end{eqnarray}
where $\phi_{ij}=\phi(\|x_i-x_j\|)~(i, j=1, ..., m)$. Then the RBF weights are determined
to be
\begin{eqnarray}
\left(
  \begin{array}{c}
    \omega_1 \\
    \omega_2 \\
    ... \\
    \omega_m \\
  \end{array}
\right)
=
\left(
  \begin{array}{cccc}
    \phi_{11} & \phi_{12} & ... & \phi_{1m} \\
    \phi_{21} & \phi_{22} & ... & \phi_{2m} \\
    ...       & ...       & ... & ... \\
    \phi_{m1} & \phi_{m2} & ... & \phi_{mm} \\
  \end{array}
\right)^{-1}
\left(
  \begin{array}{c}
    d_1 \\
    d_2 \\
    ... \\
    d_m \\
  \end{array}
\right).
\end{eqnarray}
Once the weights are obtained with the $m$ samples $(x_i, d_i)$, the reconstructed function $S(x)$
can be calculated with Eq.~(\ref{Eq:Sx}) for any point $x$.

As in Ref.~\cite{Wang2011PRC}, the Euclidean norm is defined to be the distance between nuclei $(Z_i,N_i)$ and $(Z_j,N_j)$ in nuclear chart:
\begin{eqnarray}
   r=\sqrt{(Z_i-Z_j)^2 + (N_i-N_j)^2}.
\end{eqnarray}
The basis function $\phi(r)=r$ is adopted in this work, since the mass deviation can be
reconstructed relatively better with $\phi(r)=r$ than other basis functions~\cite{Wang2011PRC}. Then the mass difference $D(Z,N)=M_{\textrm{exp}}(Z,N)-M_{\textrm{th}}(Z,N)$ between the experimental data $M_{\textrm{exp}}$ and those predicted with nuclear mass models $M_{\textrm{th}}$ could be reconstructed with Eq.~(\ref{Eq:Dx}). Once the weights are obtained, the reconstructed function $S(Z,N)$ can be calculated with Eq.~(\ref{Eq:Sx}) for any nucleus $(Z,N)$. Then the revised mass for nucleus $(Z,N)$ is given by
\begin{eqnarray}
M_{\textrm{th}}^{\textrm{RBF}}(Z,N)=M_{\textrm{th}}(Z,N)+S(Z,N).
\end{eqnarray}
For training the RBF with Eq.~(\ref{Eq:Dx}), only those nuclei between the minimum distance $R_{\textrm{min}}$ and maximum distance $R_{\textrm{max}}$ are involved, i.e. $R_{\textrm{min}} \leqslant r \leqslant R_{\textrm{max}}$. If the reconstructed function $S(Z,N)$ for a nucleus is obtained by training the RBF including itself, i.e. $R_{\textrm{min}}=0$, it is clear that $S(Z,N)$ is just the $D(Z,N)$ and hence $M_{\textrm{th}}^{\textrm{RBF}}(Z,N)=M_{\textrm{exp}}(Z,N)$. Therefore, to test the predictive power of RBF approach, the function $S(Z,N)$ for a known nucleus should be reconstructed with $R_{\textrm{min}}\geqslant 1$.

To evaluate the predictive power of RBF approach, the rms deviation, i.e.,
\begin{equation}
    \sigma_{\textrm{rms}}
   =\sqrt{\frac1n
    \sum_{i=1}^n(M_i^{\textrm{th}}-M_i^{\textrm{exp}})^2},
\end{equation}
is employed, where $M_i^{\textrm{th}}$ and $M_i^{\textrm{exp}}$ are the theoretical and
experimental nuclear masses, respectively, and $n$ is the number of nuclei contained in
a given set. In this investigation, we only consider nuclei with $N\geqslant 8$ and
$Z\geqslant 8$ and the experimental data are taken from AME12~\cite{Audi2012CPC},
unless otherwise specified. For the theoretical mass models, we take RMF~\cite{Geng2005PTP}, HFB-21~\cite{Goriely2010PRC}, DZ10~\cite{Duflo1995PRC}, DZ31~\cite{Zuker2008RMF}, ETFSI-2~\cite{Goriely2000AIP}, KTUY~\cite{Koura2005PTP}, FRDM~\cite{Moller1995ADNDT}, and
WS3~\cite{Wang2011PRC} mass models as examples, with the rms deviation spanning from $2.2$ MeV to $0.3$ MeV with respect to experimental data in AME12. For convenience, the $\sigma_{\textrm{rms}}(\textrm{Model})$ and $\sigma_{\textrm{rms}}(\textrm{Model+RBF})$ of nuclear mass models are denoted as $\sigma_{\textrm{rms0}}(\textrm{Model})$ and $\sigma_{\textrm{rmsR}}(\textrm{Model})$, hereafter.

\section{Results and discussions}
\begin{table}
\begin{center}
\caption{The rms deviations in MeV between known masses in AME12 and predictions of various nuclear mass models without (the second column) and with (the third column) the RBF approach. The fourth column gives the reduction of the rms deviations after combining the RBF approach.} \label{tb1}
\begin{tabular}{ccccccc}
\hline \hline
Model      &    &$\sigma_\textrm{rms0}$
                           &    &$\sigma_\textrm{rmsR}$
                                            &   &Reduction (\%)
                                                                \\ %Num
\hline
RMF        &    &2.217     &    &0.488      &   &78\%           \\ %2328
HFB-21     &    &0.572     &    &0.410      &   &28\%           \\ %2353
DZ10       &    &0.591     &    &0.225      &   &62\%           \\ %2353
DZ31       &    &0.397     &    &0.204      &   &49\%           \\ %2353
ETFSI-2    &    &0.719     &    &0.360      &   &50\%           \\ %2234
KTUY       &    &0.701     &    &0.210      &   &70\%           \\ %2353
FRDM       &    &0.654     &    &0.268      &   &59\%           \\ %2353
WS3        &    &0.335     &    &0.207      &   &38\%           \\ %2353
\hline \hline
\end{tabular}
\end{center}
\end{table}
For testing the predictive power of RBF approach, we first reconstruct the function $S(N,Z)$ for a known nucleus based on the remaining known masses in AME12 and the predictions in nuclear mass models. In other words, we take $R_{\textrm{min}}=1$ and $R_{\textrm{max}}=1000$ (no limits on $R_{\textrm{max}}$) for training the RBF. The corresponding results are given in Table~\ref{tb1}. It is found that the reduction of the rms deviation exceeds $25\%$ for all mass models considered here. In particular, the  largest improvement with $78\%$ reduction of rms deviation is obtained for the RMF mass model. The corresponding rms deviation is reduced from $2.2$ MeV to $0.5$ MeV, which is comparable to the corresponding rms deviation in the microscopic HFB-21 mass model. Therefore, the reduction of rms deviations clearly shows that the predictive accuracy of the nuclear mass models can be significantly improved by combining the RBF approach.

\begin{figure}[h]
\includegraphics[width=7cm]{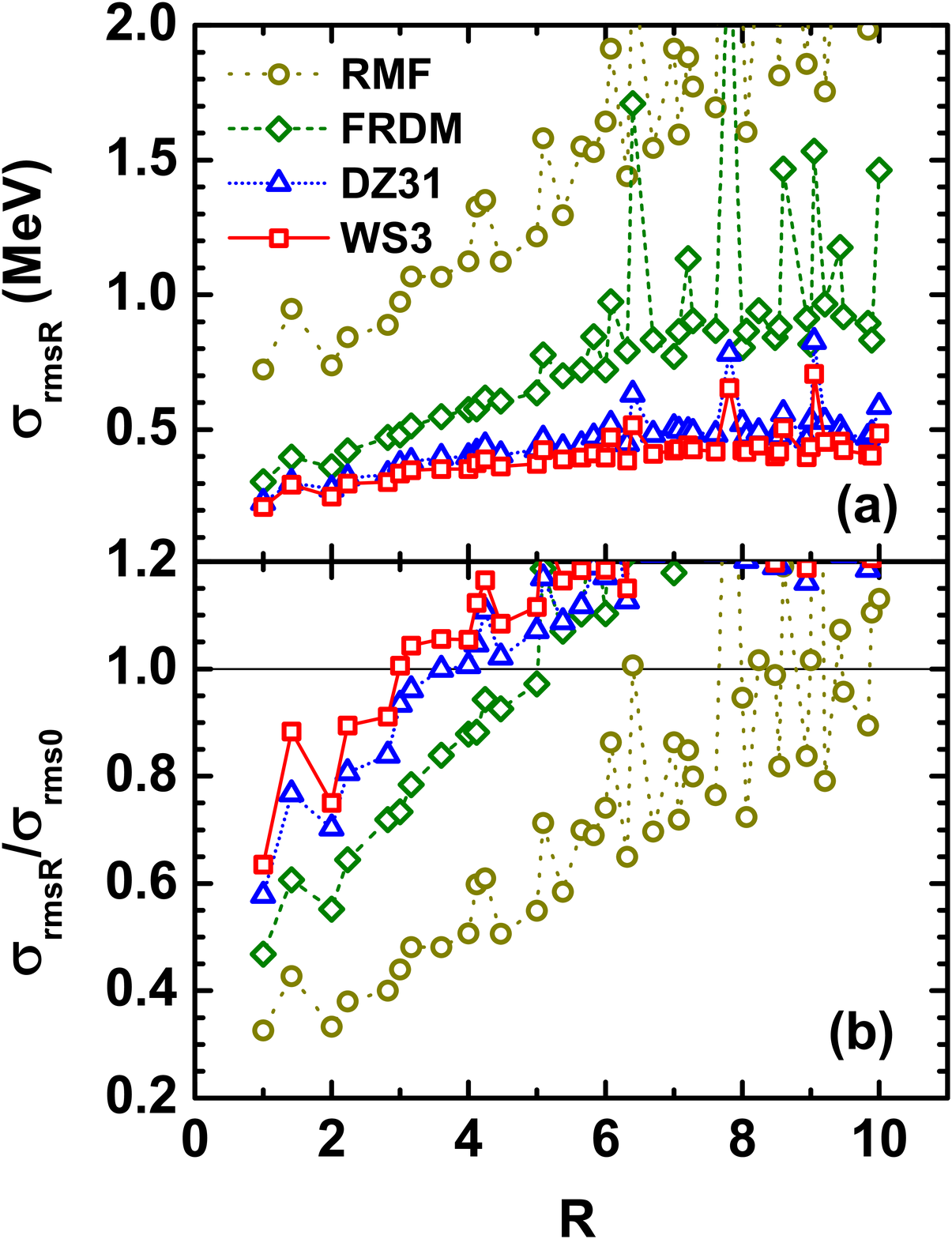}\\
\caption{(Color online) The rms deviations $\sigma_\textrm{rmsR}$ and the relative rms deviations $\sigma_\textrm{rmsR}/\sigma_\textrm{rms0}$ with respect to the known masses in AME12 for different mass models. For training the RBF for a certain nucleus, only those nuclei with $r=R$ are involved.}\label{fig1}
\end{figure}
In the calculations of Table~\ref{tb1}, the reconstructed function $S(N,Z)$ for a nucleus is obtained by training the RBF with the remaining known masses in AME12. For better understanding the predictive power of RBF, it is necessary to investigate the mass correlations between this nucleus and those nuclei used in training the RBF. In Fig.~\ref{fig1}, the rms deviations $\sigma_\textrm{rmsR}$ and the relative rms deviations $\sigma_\textrm{rmsR}/\sigma_\textrm{rms0}$ with respect to the known masses in AME12 are shown as a function of $R$, which is the distance between the selected nucleus and the nuclei used in training the RBF. For clarity, only the results of RMF, DZ31, FRDM, and WS3 mass models are shown in the figure. In fact, other mass models show similar trends and their corresponding results are almost between the results of RMF and WS3 mass models. From Fig.~\ref{fig1}, it is clear that $\sigma_\textrm{rmsR}$ generally increases as the increase of $R$, while the order of $\sigma_\textrm{rmsR}$ generally remains the same as that of $\sigma_\textrm{rms0}$. For the WS3, DZ31, and FRDM mass models, the RBF approach ceases to improve the mass predictions, i.e. $\sigma_\textrm{rmsR}/\sigma_\textrm{rms0}>1$, when the distances $R$ are larger than $3$, $4$, and $5$, respectively. However, the RBF approach can improve the mass prediction of RMF model even with some nuclei around $R=10$. This long correlation may imply some important physics correlations are missing in this RMF mass model, and hence a larger rms deviation with respect to the known masses.

\begin{figure}[h]
\includegraphics[width=7cm]{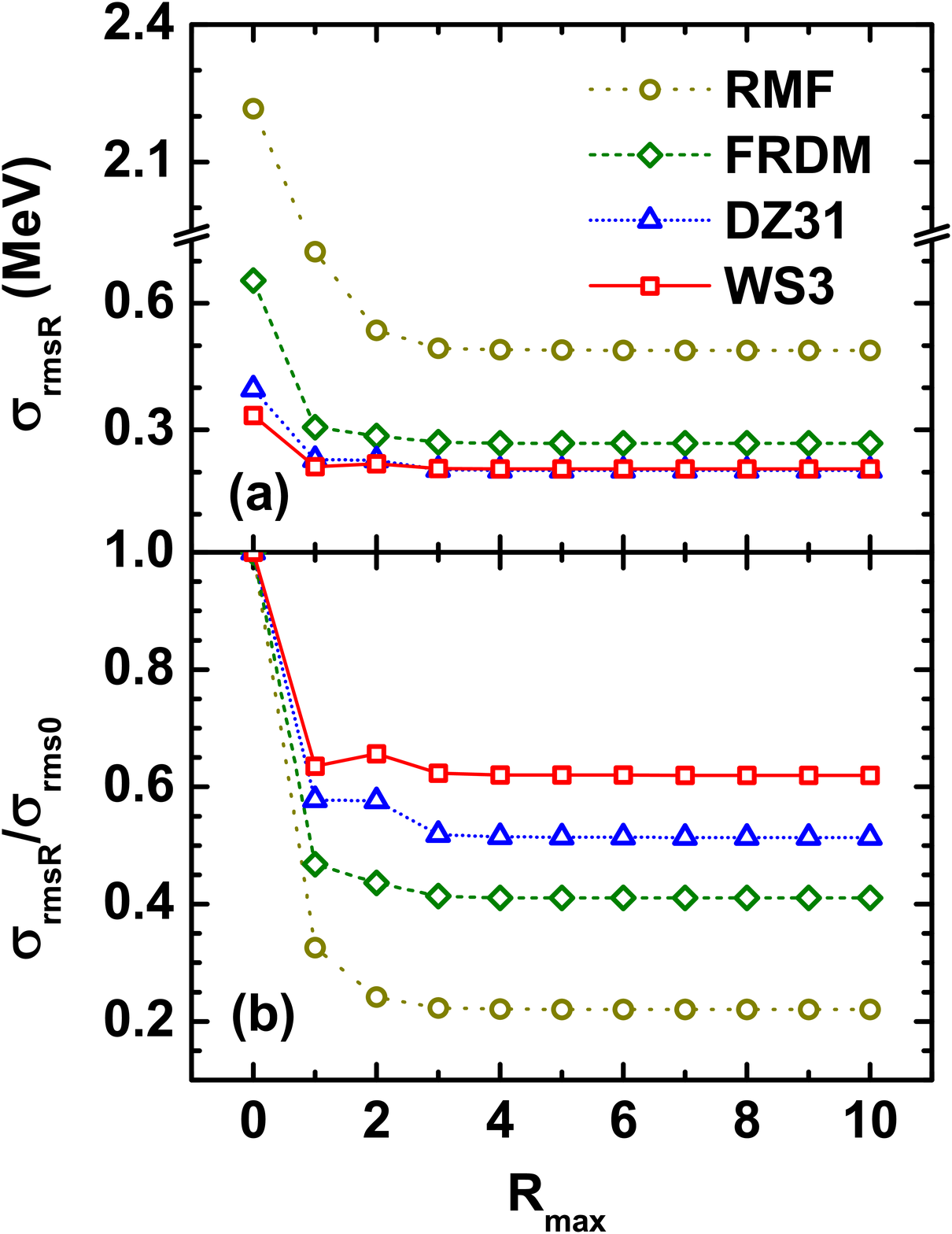}\\
\caption{(Color online) The rms deviations $\sigma_\textrm{rmsR}$ and the relative rms deviations $\sigma_\textrm{rmsR}/\sigma_\textrm{rms0}$ with respect to the known masses in AME12 for different mass models as a function of $R_{\textrm{max}}$. For training the RBF, $R_{\textrm{min}}$ is fixed to be $R_{\textrm{min}}=1$. }\label{fig2}
\end{figure}
Furthermore, we investigate the accumulative rms deviations by training the RBF with nuclei at distance between $R_{\textrm{min}}$ and $R_{\textrm{max}}$. By fixing $R_{\textrm{min}}=1$, the rms deviations $\sigma_\textrm{rmsR}$ and the relative rms deviations $\sigma_\textrm{rmsR}/\sigma_\textrm{rms0}$ as a function of $R_{\textrm{max}}$ are shown in Fig.~\ref{fig2} for different mass models. The points at $R_{\textrm{max}}=0$ mean no masses are included in training the RBF, so $\sigma_\textrm{rmsR}$ is just $\sigma_\textrm{rms0}$. From Fig.~\ref{fig2}, one can see that the nuclei at distance $r=1$ play an important role in improving the predictive accuracy of different mass models. By further including the nuclei in the range of $1< r \leqslant 3$ for training the RBF, the rms deviations can be slightly reduced. However, the improvement in the predictive accuracy is almost negligible with the inclusion of nuclei at $r>3$ for all mass models, although the nuclei at $r>3$ can still help the RBF approach improve the mass predictions for some mass models, such as the FRDM and RMF models. This indicates that the RBF approach can well extract most mass correlations only from those nuclei with $r \leqslant 3$.

\begin{figure}[h]
\includegraphics[width=7cm]{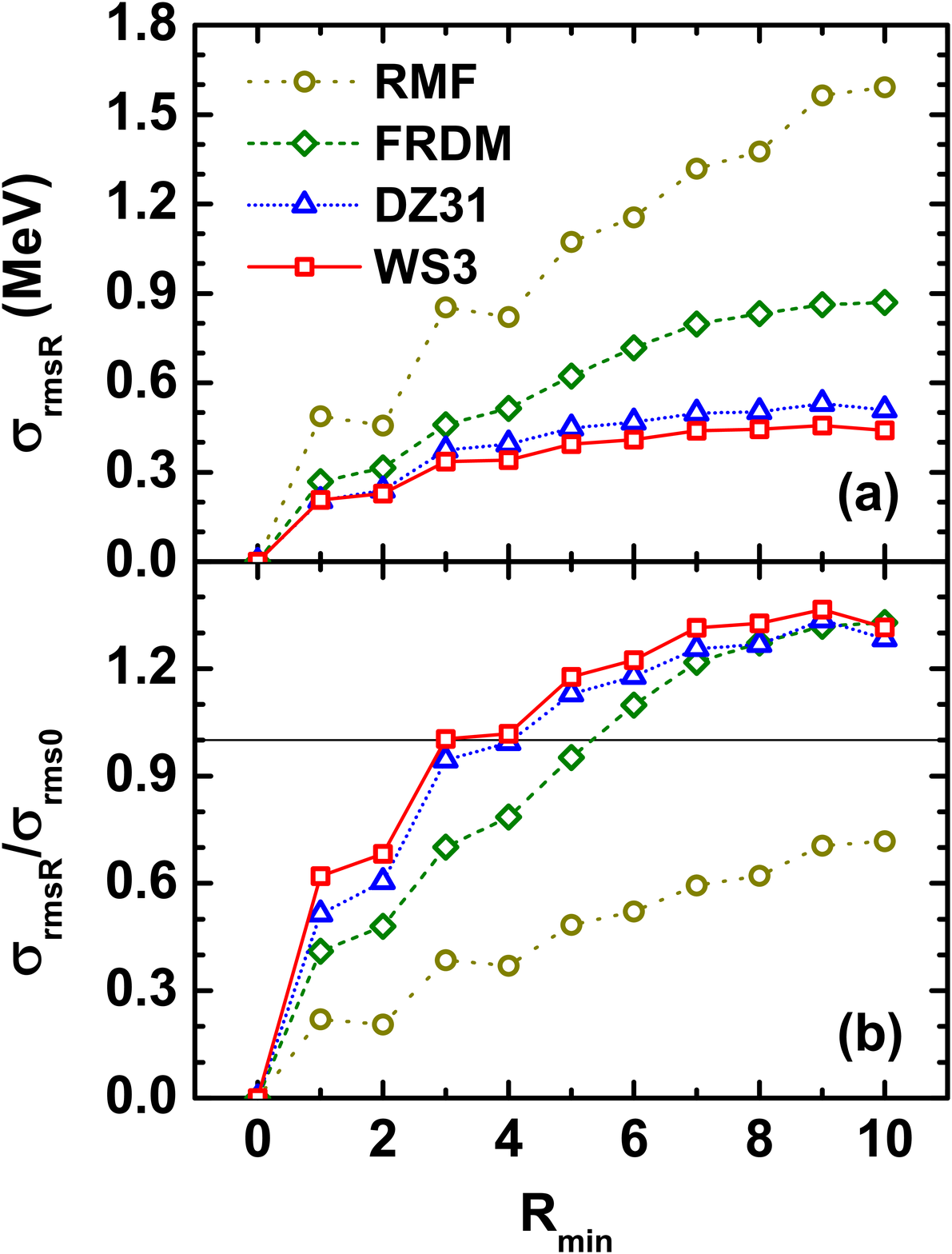}\\
\caption{(Color online) The rms deviations $\sigma_\textrm{rmsR}$ and the relative rms deviations $\sigma_\textrm{rmsR}/\sigma_\textrm{rms0}$ with respect to the known masses in AME12 for different mass models as a function of $R_{\textrm{min}}$. For training the RBF, there are no limits on $R_{\textrm{max}}$.}\label{fig3}
\end{figure}
On the other hand, by fixing the $R_{\textrm{max}}=1000$ (no limits on $R_{\textrm{max}}$), the rms deviations $\sigma_\textrm{rmsR}$ and the relative rms deviations $\sigma_\textrm{rmsR}/\sigma_\textrm{rms0}$ as a function of $R_{\textrm{min}}$ are shown in Fig.~\ref{fig3} for different mass models. The points at $R_{\textrm{min}}=0$ mean the reconstructed function $S(Z,N)$ for one nucleus is obtained by training the RBF including itself, so $M_{\textrm{th}}^{\textrm{RBF}}=M_{\textrm{exp}}$ and hence $\sigma_\textrm{rmsR}=0$. The points at $R_{\textrm{min}}=1$ just correspond to those $\sigma_\textrm{rmsR}$ in Table~\ref{tb1}. If the nuclei with $r=1$ are excluded from training RBF, the rms deviation $\sigma_\textrm{rmsR}$ increases for the DZ31, FRDM, and WS3 mass models, while it decreases for the RMF mass model. However, the influence on $\sigma_\textrm{rmsR}$ is unremarkable, so the RBF approach can also remarkably improve the mass predictive accuracy only with the nuclei of $r\geqslant 2$. Furthermore, if we exclude the nuclei with $r \leqslant 2$, $\sigma_\textrm{rmsR}$ is systematically increased for all mass models. For mass models with smaller $\sigma_\textrm{rms0}$, i.e. DZ31, and WS3 models, the rms deviations $\sigma_\textrm{rmsR}$ at $R_{\textrm{min}}=3$ are similar to $\sigma_\textrm{rms0}$, which means the RBF approach ceases to improve the model predictions effectively. However, the RBF approach is still effective for the FRDM and RMF mass models, even if the nuclei with $r \leqslant 3$ are excluded. From Fig.~\ref{fig1}, it is known that these two mass models have a relatively longer mass correlations even with the nuclei at $r>3$, so RBF approach still remarkably reduce their model deviations. However, it should be noted that the order of $\sigma_\textrm{rmsR}$ among different models almost remains unchanged at various $R_{\textrm{min}}$ in Fig.~\ref{fig3}, i.e. $\sigma_\textrm{rmsR}$(WS3)$<\sigma_\textrm{rmsR}$(DZ31)$<\sigma_\textrm{rmsR}$(FRDM)$<\sigma_\textrm{rmsR}$(RMF).

\begin{figure}[h]
\includegraphics[width=7.5cm]{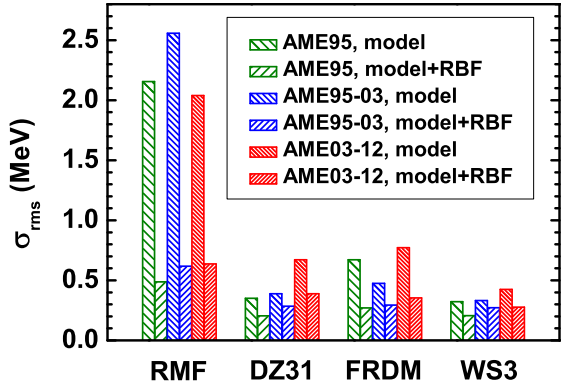}\\
\caption{(Color online) The rms deviations $\sigma_\textrm{rms0}$ and $\sigma_\textrm{rmsR}$ with respect to the known masses for different mass models in the AME95-03-12 test. For training the RBF, only those nuclei in AME95 are employed, while the their masses are taken from AME12.}\label{fig4}
\end{figure}

\begin{figure}[h]
\includegraphics[width=8.5cm]{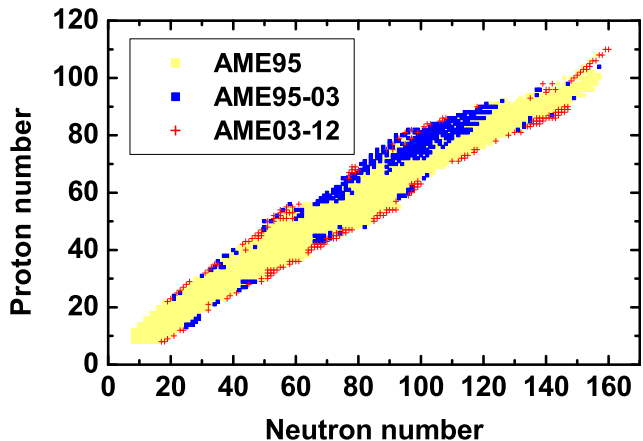}\\
\caption{(Color online) Positions of nuclei in the AME95-03-12 test.}\label{fig5}
\end{figure}
The AME95-03 has been extensively employed to check the predictive power of nuclear mass models in the literatures~\cite{Wang2011PRC, Hua2012SCMPA} and we extend this test to the AME95-03-12 in this work. In the AME95-03-12 test, the nuclei in AME12 are separated into three subsets, i.e.,
the $1758$ nuclei in atomic mass evaluation of 1995 (AME95)~\cite{Audi1995NPA}, the $381$ nuclei first appearing in AME03~\cite{Audi2003NPA}, and the $214$ ``new" nuclei appearing in AME12. The rms deviations $\sigma_\textrm{rms0}$ and $\sigma_\textrm{rmsR}$ for different mass models in the AME95-03-12 test are shown in Fig.~\ref{fig4}. For training the RBF, only those nuclei in AME95 are employed, while their masses are taken from AME12. From this figure, it is clear that the RBF approach significantly reduces the rms deviations of different mass models, especially for those models with large $\sigma_\textrm{rms0}$. The rms deviations $\sigma_\textrm{rmsR}$ with respect to the masses of 1758 nuclei in AME95 are all within $0.5$ MeV and the best predictive accuracy can reduce to $0.206$ MeV based on the WS3 mass models. The rms deviations $\sigma_\textrm{rmsR}$ with respect to the masses of $381$ nuclei first appearing in AME03 are slightly increased, while they are still remarkably smaller than the corresponding $\sigma_\textrm{rms0}$. The $214$ ``new" masses appearing in AME12 are not used in determining the effective interactions of all mass models considered here, so these new data in AME12 are worthwhile to test the predictive power of RBF approach. It is found that the rms deviations $\sigma_\textrm{rmsR}$ with respect to the $214$ ``new" masses appearing in AME12 are also remarkably reduced, and the best predictive accuracy is $0.277$ MeV based on the WS3 mass models. From Fig.~\ref{fig5}, it is shown that the $381$ nuclei first appearing in AME03 and the $214$ ``new" nuclei appearing in AME12 are mainly around the nuclei in AME95 with $r\lesssim 3$, so the RBF approach can remarkably reduce the rms deviations for these nuclei.

\begin{table}
\begin{center}
\caption{The rms deviations in MeV with respect to the known masses in AME12 for the nuclei, whose masses are evaluated values in AME03 (marked by ``\#" in the mass table of AME03). The second column represents the rms deviations $\sigma_\textrm{rms0}$ for different mass models. The third and fourth columns both represent the rms deviations $\sigma_\textrm{rmsR}$ by training the RBF with the nuclei in AME03, while the data are taken from AME03 and AME12, respectively.} \label{tb2}
\begin{tabular}{ccccccc}
\hline \hline
Model      &    &$\sigma_\textrm{rms0}$
                            &    &$\sigma_\textrm{rmsR}$(AME03)
                                              &    &$\sigma_\textrm{rmsR}$(AME12)\\
\hline
RMF        &    &1.956      &    &0.604       &    &0.620      \\
HFB-21     &    &0.631      &    &0.559       &    &0.483      \\
DZ10       &    &0.876      &    &0.419       &    &0.327      \\
DZ31       &    &0.673      &    &0.430       &    &0.325      \\
ETFSI-2    &    &0.690      &    &0.484       &    &0.415      \\
KTUY       &    &1.102      &    &0.415       &    &0.325      \\
FRDM       &    &0.771      &    &0.469       &    &0.371      \\
WS3        &    &0.425      &    &0.375       &    &0.268      \\
\hline \hline
\end{tabular}
\end{center}
\end{table}

\begin{figure}[h]
\includegraphics[width=7.5cm]{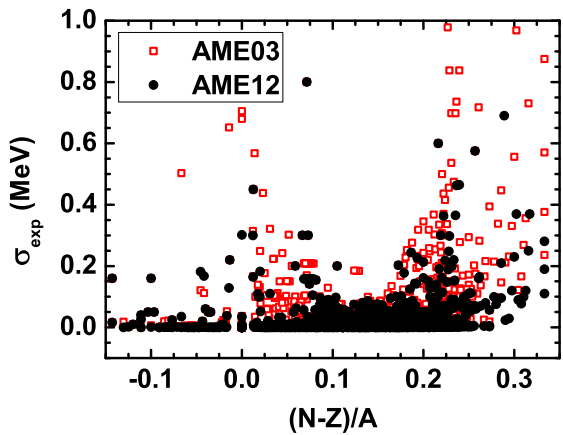}\\
\caption{(Color online) The experimental uncertainties as a function of the isospin asymmetry $I=(N-Z)/A$. The squares and circles represent the experimental uncertainties in AME03 and AME12, respectively.}\label{fig6}
\end{figure}
By using the systematic trends in the mass surface and its derivative, the mass evaluation method in AME provides the best short-range mass extrapolation~\cite{Lunney2003RMP}. Therefore, it is interesting to compare the accuracy between the RBF approach and the method used in AME. Taking the nuclei whose masses are evaluated values in AME03 (marked by ``\#" in the mass table of AME03) as an example, the rms deviations with respect to the new experimental data in AME12 are $0.398$ MeV. For comparison with the method in AME at the same foot, the experimental data employed in training the RBF should be taken from AME03 as well, and the corresponding results based on different mass models are given in the third column of Table~\ref{tb2}. In addition, the rms deviations $\sigma_\textrm{rms0}$ for these nuclei are given in the second column of Table~\ref{tb2}. Clearly, the RBF approach also remarkably improves the predictive power of various mass models. It should be pointed out that the rms deviation $\sigma_\textrm{rmsR}$ based on the WS3 mass model is even smaller than that from the method in AME. In Fig.~\ref{fig6}, the experimental uncertainties in AME12 and AME03 are presented as a function of the isospin asymmetry $I=(N-Z)/A$. It is found that the experimental uncertainties are significantly improved in AME12 comparing with those in AME03, especially for those nuclei around the border region of experimental data with $I\lesssim 0.1$ and $I\gtrsim 0.2$. Therefore, we further update the data in training the RBF with those in AME12 and the corresponding results are shown in the fourth column of Table~\ref{tb2}. It is clear that $\sigma_\textrm{rmsR}$ can be significantly reduced for most mass models. Based on the WS3 mass model, the predictive accuracy has been reduced to $0.268$ MeV. Therefore, the new high-precision experimental data are also very important to improve the nuclear mass models with RBF approach.

\begin{figure}[h]
\includegraphics[width=8.5cm]{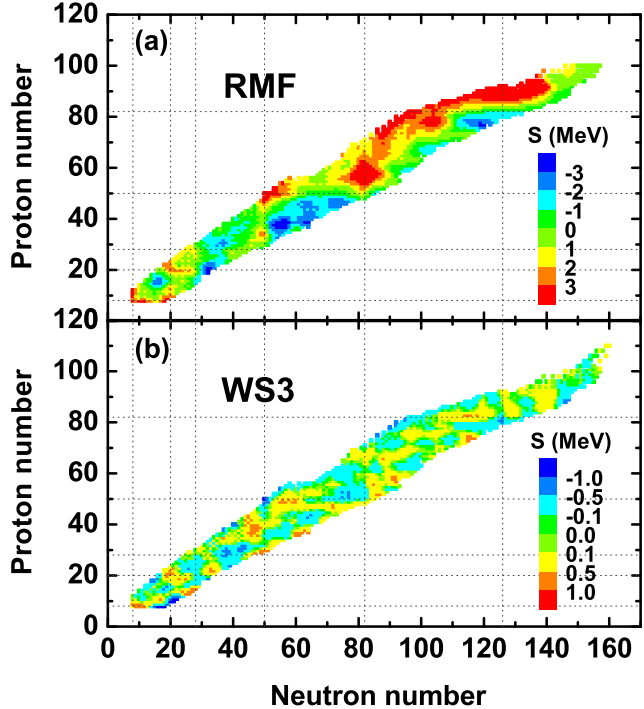}\\
\caption{(Color online) The reconstructed functions $S(Z,N)$ based on the measured masses in the AME12 for the RMF [panel (a)] and WS3 [panel (b)] mass models. The dotted lines denote the magic numbers.}\label{fig7}
\end{figure}

\begin{figure}[h]
\includegraphics[width=8.5cm]{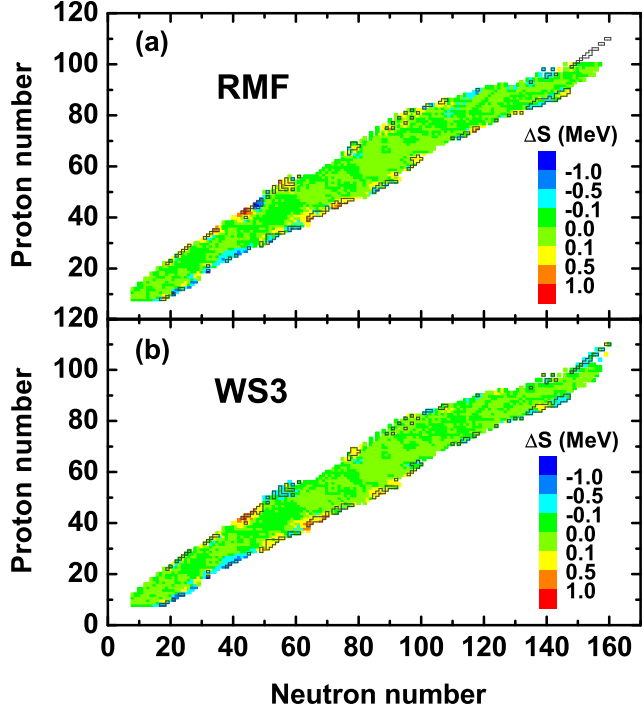}\\
\caption{(Color online) The differences of the reconstructed functions $S(N,Z)$ between those based on the measured masses in the AME12 and those based on the measured masses in AME03. The panels (a) and (b) correspond to the RMF and WS mass models, respectively. The ``new" nuclei in AME12 are indicated by the black contours.}\label{fig8}
\end{figure}
To investigate the radial basis function corrections in detail, the reconstructed functions $S(Z,N)$ based on the measured masses in the AME12 are shown in Fig.~\ref{fig7} by taking the RMF and WS3 mass models as examples. It is clear that the reconstructed functions are sensitive to the the nuclear mass models. For the RMF mass model, the $S(Z,N)$ are about $3$ MeV for the nuclei near $(Z, N)=(50, 50)$, $(Z, N)=(58, 82)$, $(Z, N)=(78, 92)$, and $(Z, N)=(92, 126)$. This just corresponds to the nuclei whose masses are underestimated in the RMF model~\cite{Geng2005PTP}. Moreover, the overestimation of nuclear masses in the regions near $(Z, N)=(38, 60)$ and $(Z, N)=(78, 120)$ in RMF model are also well improved by the RBF approach with the $S(Z,N)\sim-3$ MeV in these two regions. For the WS3 mass model, it better describes the nuclear masses than the RMF mass model, while there still exists small but systematically correlated errors~\cite{Liu2011PRC, Wang2013JPCS}. With the RBF approach, these systematic correlations can be well extracted as well. To further investigate the influence of the ``new" masses in AME12 on improving the nuclear mass models with the RBF approach, the differences of the reconstructed functions $S(N,Z)$ between those based on the measured masses in the AME12 and those based on the measured masses in AME03 are shown in Fig.~\ref{fig8}. It is found that the differences of reconstructed functions $S(N,Z)$ are generally within $100$ keV for most nuclei, while it is relatively larger for those nuclei around the border region of experimental data, especially for those ``new" nuclei in AME12. This can be well understood since significant improvements in the mass measurements are made for nuclei near the border region in recent years~\cite{Audi2012CPC}. Therefore, the RBF approach is sensitive to the experimental masses and it is necessary to adopt the high-precision experimental data to improve the nuclear mass models.

\section{Summary and perspective}
In this work, the mass correlations in the radial basis function (RBF) approach are carefully investigated based on $8$ widely used nuclear mass models, ranging from macroscopic-microscopic to microscopic types. The mass correlations usually exist between a nucleus and its surrounding nuclei with distance $r\lesssim 3$. However, the correlation distance is dependent on the nuclear mass models, which can go up to the distance of $r\sim 10$ for the mass models with larger rms deviations, such as the RMF model. To extract these mass correlations, it is shown that the nuclei at distance $r\leqslant 3$ are necessary to include in the training of RBF approach. In this way, the RBF approach can make significant improvements in the mass predictions for different mass models. The AME95-03-12 test further shows that the RBF approach provides a very effective tool to improve mass predictions significantly in regions not far from known nuclear masses. Based on the latest Weizs\"{a}cker-Skyrme mass model, the RBF approach can achieve an accuracy comparable with the extrapolation method used in atomic mass evaluation, which can be further improved by the incorporation of new measurements. As claimed in the introduction, the effective interaction PC-PK1 remarkably improves the mass prediction of the RMF model. Therefore, it is interesting to investigate the predictive power of PC-PK1 mass model with the help of RBF approach when the calculated masses with PC-PK1 for all nuclei in AME12 are available in the future. In addition, considering the success in improving the nuclear mass predictions, the RBF approach has a great potential to improve theoretical calculations of other physical quantities, such as nuclear $\beta$-decay half-lives, fission barriers, and excitation spectra.

%-------------------------------------------------------------------------------------------------------
\section{Acknowledgements}
This work was partly supported by the National Natural Science Foundation of China (Grants No. 11205004, No. 11235002, No. 11175001, No. 11105010, No. 11128510, and No. 11035007), the 211 Project of Anhui University under Grant No. 02303319-33190135, the Program for New Century Excellent Talents in University of China under Grant No. NCET-09-0031, the Key Research Foundation of Education Ministry of Anhui Province of China under Grant No. KJ2012A021, and the Natural Science Foundation of Anhui Province under Grant No. 11040606M07.
%-------------------------------------------------------------------------------------------------------

%-------------------------------------------------------------------------------------------------------

%-------------------------------------------------------------------------------------------------------

%-------------------------------------------------------------------------------------------------------
\end{CJK*}
\end{document}